\documentclass[a4paper,10pt]{article}
\usepackage[margin=4cm]{geometry}
\usepackage[noblocks]{authblk}

\usepackage{setspace}
\usepackage[fleqn]{amsmath}
\usepackage{amssymb}
\usepackage{natbib}
\usepackage{titlesec}
\titleformat{\section}{\bf \uppercase  }{\thetitle}{1em}{}{}
\titleformat{\subsection}{\bf}{\thetitle}{1em}{}{}
\titleformat{\subsubsection}{\it}{\thetitle}{1em}{}{}

\titlespacing*{\section}{0pt}{6pt}{0pt}
\titlespacing*{\subsection}{0pt}{3pt}{0pt}
\titlespacing*{\subsubsection}{0pt}{3pt}{0pt}
\DeclareMathOperator*{\infimum}{inf}
\DeclareMathOperator{\trace}{Tr}

\newcommand{\prob}[1]{\mathcal{#1}}
\title{Emerging Directions in Geophysical Inversion}
\author{Andrew P.~Valentine\footnote{E-mail: andrew.valentine@durham.ac.uk.}}
\affil{Department of Earth Sciences, Durham University, South Road, Durham, DH1 3LE, UK.}
\author{Malcolm Sambridge}

\affil{Research School of Earth Sciences, The Australian National University, 142 Mills Road, Acton ACT 2601, Australia.}
\date{}

\usepackage{graphicx}
\graphicspath{{figures}}
\newcommand{\indexterm}{} 
\begin{document}
\maketitle
\section*{Abstract}
In this chapter, we survey some recent developments in the field of geophysical inversion. We aim to provide an accessible general introduction to the breadth of current research, rather than focussing in depth on particular topics. We hope to give the reader an appreciation for the similarities and connections between different approaches, and their relative strengths and weaknesses.

\section{Introduction}
Geophysics is built upon indirect information. We cannot travel deep into the Earth to directly measure rheological properties, nor journey back through geological time to record the planet's tectonic evolution. Instead, we must draw inferences from whatever observations we can make, constrained as we are to the Earth's surface and the present day. Inevitably, such datasets are sparse, incomplete, and contaminated with signals from many unknown events and processes. We therefore rely on a variety of mathematical, statistical and computational techniques designed to help us learn from available data. Collectively, these are the tools of `geophysical inversion', and they lie at the heart of all progress in geophysics.

To achieve this progress, geophysicists have long pioneered---and indeed driven---developments in the mathematical and statistical theory that underpins inference. The acclaimed French mathematician Pierre-Simon Laplace played a central role in our understanding of tidal forcing, developing the theory of spherical harmonics along the way. He is also credited (along with Gauss and Legendre) with the development of the least-squares algorithm and the underpinnings of modern Bayesian statistics---an approach which was subsequently extended and popularised within the physical sciences by Sir Harold \citet{Jeffreys1931,Jeffreys1939}, who is of course also well-known for his contributions to seismology and solid-earth geophysics \citep[see, e.g.][]{Cook1990}.  Technological developments have also been significant, with (for example) the challenges of handling and processing the huge volumes of data obtained from continuously-operating terrestrial and satellite sensor systems stimulating innovation in computational science.

In this chapter, we discuss some current and emerging ideas that we believe to have significance for the broad field of geophysical inversion. In doing so, we aim to not just highlight novelty, but also demonstrate how such `new' ideas can be connected into the canon of established techniques and methods. We hope that this can help provide insight into the potential strengths and weaknesses of different strategies, and support the interpretation and integration of results obtained using different approaches. Inevitably, constraints of time and space mean that our discussion here remains far from comprehensive; much interesting and important work must be omitted, and our account is undoubtedly biased by our own perspectives and interests. Nevertheless, we hope that the reader is able to gain some appreciation for the current state of progress in geophysical inversion.

In order to frame our discussion, and to enable us to clearly define notation and terminology, we begin with a brief account of the basic concepts of geophysical inversion. For a more in-depth account, readers are encouraged to refer to one of the many textbooks and monographs covering the subject, such as those by \citet{Menke1989}, \citet{Parker1994}, \citet{Tarantola2005} or \citet{Aster2013}.

\section{Fundamentals}
The starting point for any geophysical inversion must be a mathematical description of the earth system of interest. In practical terms, this amounts to specifying some relationship of the form
\begin{equation}
\mathcal{F}[m(\mathbf{x},t),u(\mathbf{x},t)]=0\label{eq:genfwd}
\end{equation}
where $m(\mathbf{x},t)$ represents some property (or collection of properties) of the Earth with unknown value that may vary across space, $\mathbf{x}$, and/or time, $t$; and where $u(\mathbf{x},t)$ represents some quantity (or collection of quantities) that can---at least in principle---be measured or observed. Most commonly in geophysics, $\mathcal{F}$ has the form of an integro-differential operator.  Underpinning eq.~(\ref{eq:genfwd}) will be some set of assumptions, $\mathcal{A}$, although these may not always be clearly or completely enunciated. 

\subsection{The Forward Problem}
The fundamental physical theory embodied by eq.~(\ref{eq:genfwd}) may then be used to develop predictions, often via a computational simulation. This invariably involves introducing additional assumptions, $\mathcal{B}$. In particular, it is common to place restrictions on the function $m$, so that it may be assumed to have properties amenable to efficient computation. For example, it is very common to assert that the function must lie within the span of a finite set of basis functions, $\psi_{1},\ldots,\psi_M$, allowing it to be fully-represented by a set of $M$ expansion coefficients,
\begin{equation}
m(\mathbf{x},t) = \sum_{i=1}^M m_{i}\psi_i(\mathbf{x},t)\label{eq:basis}
\end{equation}
It is important to recognise that such restrictions are primarily motivated by computational considerations, but may impose certain characteristics---such as a minimum length-scale, or smoothness properties---upon the physical systems that can be represented. Nevertheless, by doing so, we enable eq.~(\ref{eq:genfwd}) to be expressed, and implemented, as a `forward model'
\begin{equation}
u(\mathbf{x},t) = \mathcal{G}(\mathbf{x},t,m)\label{eq:fwdprob}
\end{equation}
which computes simulated observables for any `input model' conforming to the requisite assumptions. Typically, the function $\mathcal{G}$ exists only in the form of a numerical computer code, and not as an analytical expression in any meaningful sense. As a result, we often have little concrete understanding of the function's global behaviour or properties, and the computational cost associated with each function evaluation may be high.

\subsection{Observational Data and the Inverse Problem }
We use $\mathbf{d}$ to represent a data vector, with each element $d_i$ representing an observation made at a known location in space and time, $(\mathbf{x}_i,t_i)$. This is assumed to correspond to $u(\mathbf{x}_i,t_i)$, corrupted by `noise' (essentially all processes not captured within our modelling assumptions, $\mathcal{A}\cup\mathcal{B}$), any limitations of the measurement system itself, and any preprocessing (e.g. filtering) that has been applied to the dataset. We address the latter two factors by applying transformations (e.g. equivalent preprocessing and filters designed to mimic instrument responses) to the output of our forward model; mathematically, this amounts to composing $\mathcal{G}$ with some transfer function $\mathcal{T}$. For notational convenience, we define a new function, $\mathbf{g}$, which synthesises the entire dataset $\mathbf{d}$: $\left[\mathbf{g}(m)\right]_i = \mathcal{T}\circ\mathcal{G}(\mathbf{x}_i,t_i,m)$. We also introduce the concept of a data covariance matrix, $\mathbf{C_d}$, which encapsulates our assumptions about the uncertainties and covariances within the dataset. The fundamental goal of inversion is then to find---or somehow characterise---$m$ such that $\mathbf{g}(m)$ matches or explains $\mathbf{d}$.

Since $\mathbf{d}$ contains noise, we do not expect any model to be able to reproduce the data perfectly. Moreover, the forward problem may be fundamentally non-unique: it may generate identical predictions for two distinct models. As such, there will typically be a range of models that could be taken to `agree with' observations. We must therefore make a fundamental decision regarding the approach we wish to take. We may:
\begin{enumerate}
\item Seek a single model, chosen to yield predictions that are `as close as possible' to the data, usually with additional requirements  that impose characteristics we deem desirable and ensure that a unique solution exists to be found, e.g. that the model be `as smooth as possible';
\item Seek a collection or ensemble of models, chosen to represent the spectrum of possibilities that are compatible with observations---again, perhaps tempered by additional preferences;
\item Disavow the idea of recovering a complete model, and instead focus on identifying specific characteristics or properties that must be satisfied by any plausible model.
\end{enumerate}
In the context of this paper, we have deliberately framed these three categories to be quite general in scope. Nevertheless, readers may appreciate some specific examples: the first category includes methods based upon numerical optimisation of an objective function, including the familiar least-squares algorithm \citep[e.g.][]{Nocedal1999}, while Markov chain Monte Carlo and other Bayesian methods fall within the second \citep[e.g.][]{Sambridge2002}; Backus-Gilbert theory \citep[e.g.][]{Backus1968} lies within the third. Each of these groups is quite distinct---at least in philosophy---from the others, and in the remainder of this paper we address each in turn.

\section{Single Models}\label{sec:single}
\begin{figure}
\includegraphics[width=\textwidth]{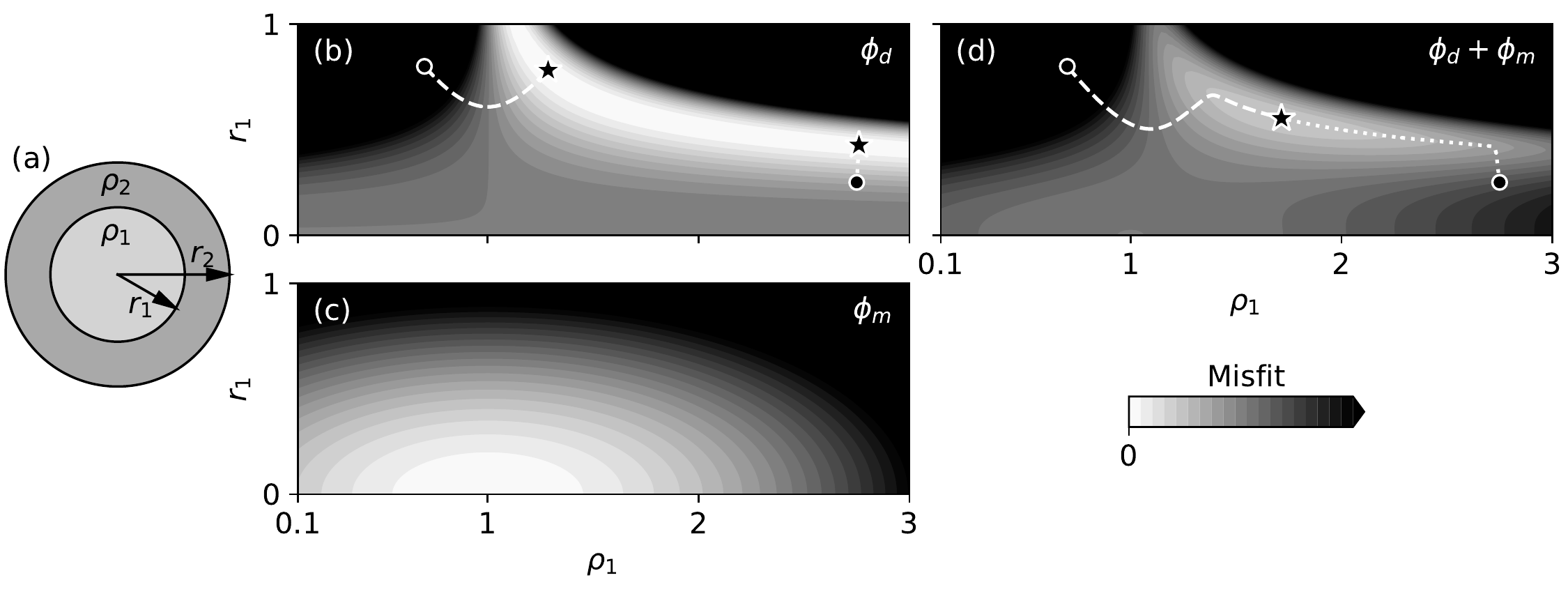}
\caption{\label{fig:deterministic}Misfit functions for a simple inverse problem \citep[after][]{Valentine2020}. (a) A planet is modelled as comprising two spherical layers: a core of radius $r_1$ and density $\rho_1$, and an outer unit of density $\rho_2$ extending to radius $r_2$. Defining units such that $r_2=1$ and $\rho_2=1$, we find the overall mass of the planet to be $M=4.76\pm0.25$ units. What can be said about $r_1$ and $\rho_1$? (b) The data misfit, $\phi_d(d,g(r_1,\rho_1))$, as in eq.~(\ref{eq:euclid}), highlighting non-linear behaviours. Two gradient-based optimisation trajectories are shown for different starting points (circles), with convergence to distinct solutions (stars). The inverse problem is inherently non-unique. (c) Penalty term, $\phi_m(r_1,\rho_1)$, expressing a preference for a small core with density similar to that of the surface layer. (d) Combined (regularised) misfit, $\phi_d(d,g(r_1,\rho_1))+\phi_m(r_1,\rho_1)$. Both optimisation trajectories now converge to the same point.}
\end{figure}

Before we can set out to find the model that `best' explains the data, we must introduce some measure of the agreement between observations and predictions. This `\indexterm{misfit function}' or `objective function' is of fundamental importance in determining the properties of the recovered model, and the efficiency of the solution algorithms that may be available to us. In general, misfit functions take the form
\begin{equation}
\phi(m) = \phi_d(\mathbf{d},\mathbf{g}({m}))+\phi_m({m})
\end{equation}
where $\phi_d$ is a metric defined in the `data space', measuring how far a model's predictions are from observations, and $\phi_m$ is a `regularisation' or `penalty' term (see Fig.~\ref{fig:deterministic}). This encapsulates any preferences we may have regarding the solution, and aims to ensure that the function $\phi$ has a unique minimum. 

Once a misfit function has been defined, it is conceptually straightforward to search for the model that minimises $\phi(m)$. However, it is often challenging to achieve this in practice. The most complete characterisation of $\phi$ comes from a \indexterm{grid-search} strategy, with systematic evaluation of the function throughout a discretised `model space' (typically following eq.~\ref{eq:basis}). This is viable for small problems, and is commonly-encountered in the geophysical literature \citep[e.g.][]{Sambridge1986,Dinh2019,Hejrani2020}, but the computational costs of evaluating the forward model, combined with the `curse of dimensionality' \citep{Curtis2001,Fernandez-Martinez2020} rapidly become prohibitive. However, in many cases, it is possible to obtain Fr\'echet derivatives of the forward problem (eq.~\ref{eq:fwdprob}) with respect to the model, $\delta \mathcal{G}/\delta m$, and this information can be used to guide a search towards the minimum of $\phi(m)$.

\subsection{Euclidean Data Metrics}\label{sec:euclid}

Overwhelmingly, the conventional choice for $\phi_d$ is the squared $L_2$, or Euclidean, norm of the residuals weighted using the data covariance matrix, $\mathbf{C_d}$,
\begin{equation}\label{eq:euclid}
\phi_d(\mathbf{d},\mathbf{g}(m)) = \bigl\|\mathbf{C_d^{-\frac{1}{2}}}\left(\mathbf{d}-\mathbf{g}(m)\right)\bigr\|_2^2 = \left(\mathbf{d}-\mathbf{g}(m)\right)^\mathbf{T}\mathbf{C_d^{-1}}\left(\mathbf{d}-\mathbf{g}(m)\right)
\end{equation}
Relying on the Fr\'echet derivatives is essentially an assumption that $g(m)$ is (locally) linear. For the usual case, where the model has been discretised as in eq.~(\ref{eq:basis}) and can be represented as a vector of coefficients, $\mathbf{m}$, we have $\mathbf{g}(\mathbf{m}) = \mathbf{g}(\mathbf{m_0}) + \mathbf{G}(\mathbf{m}-\mathbf{m_0})$, where $\mathbf{m_0}$ is the {linearisation} point and $[\mathbf{G}]_{ij} = \left.\partial[\mathbf{g(m)}]_i/\partial m_j\right|_{\mathbf{m}=\mathbf{m_0}}$. We therefore find
\begin{equation}\label{eq:gradphi}
\frac{\partial \phi}{\partial \mathbf{m}} = 2\mathbf{G^TC_d^{-1}}\left[\mathbf{G}(\mathbf{m-m_0})-(\mathbf{d} - \mathbf{g(m_0)})\right]+\frac{\partial \phi_m}{\partial \mathbf{m}}
\end{equation}
This can be used to define an update to the model, following a range of different strategies. Setting $\mathbf{m}=\mathbf{m_0}$, we obtain the gradient of $\phi$ with respect to each coordinate direction at the point of linearisation: this information may then be used to take a step towards the optimum, using techniques such as conjugate-gradient methods \citep[as in ][]{Bozdag2016} or the L-BFGS algorithm of \citet{Liu1989}, as employed by \citet{Lei2020}. Alternatively, we can exploit the fact that at the optimum, the gradient should be zero: for a suitable choice of $\phi_m$, it is possible to solve eq.~(\ref{eq:gradphi}) directly for the $\mathbf{m}$ that should minimise the misfit within the linearised regime. This is `the' \indexterm{least-squares algorithm}, employed by many studies \citep[e.g.][]{Wiggins1972, Dziewonski1981, Woodhouse1984}. Few interesting problems are truly linear, and so it is usually necessary to adopt an iterative approach, computing a new linear approximation at each step.
\subsubsection{Stochastic Algorithms}
Since the fundamental task of optimising an objective function is also central to modern machine learning efforts, recent geophysical studies have also sought to exploit advances from that sphere.  In particular, methods based on `stochastic gradient descent' have attracted some attention \citep[e.g.][]{Herwaarden2020,Bernal-Romero2021}. These exploit the intuitive idea that the gradient obtained using all available data can be approximated by a gradient obtained using only a subset of the dataset---and that by using different randomly-chosen subsets on successive iterations of gradient descent, one may reach a point close to the overall optimum. In appropriate problems, this can yield a substantial reduction in the overall computational effort expended on gradient calculations. It should be noted that the success of this approach relies on constructing approximate gradients that are, on average, unbiased; as discussed in \citet{Valentine2016}, approximations that induce systematic errors into the gradient operator will lead to erroneous results.

\newcommand{\nb}[1]{\vspace{6pt}\begin{center}\fbox{\parbox{0.6\textwidth}{\textbf{Note:} #1}}\end{center}}

\subsection{\indexterm{Sparsity}}
As has been discussed, we commonly assume that a model can be discretised in terms of some finite set of basis functions. Usually, these are chosen for computational convenience, and inevitably there will be features in the real earth system that cannot be represented within our chosen basis. This leads to the problem of `spectral leakage' \citep{Trampert1996}: features that are unrepresentable create artefacts within the recovered model. 

In digital signal processing, the conditions for complete and accurate recovery of a signal are well-known. According to Nyquist's theorem, the signal must be band-limited and sampled at a rate at least twice that of the highest frequency component present \citep{Nyquist1928}. Failure to observe this leads to spurious features in the reconstructed signal, known as aliasing---essentially the same issue as spectral leakage. This has far-reaching consequences, heavily influencing instrument design, data collection, and subsequent processing and analysis. 

However, recent work has led to the concept of `compressed sensing' \citep{Donoho2006,Candes2008}. Most real-world signals are, in some sense, sparse: when expanded in terms of an appropriately-chosen basis (as per eq.~\ref{eq:basis}), only a few non-zero coefficients are required. If data is collected by random sampling, and in a manner designed to be incoherent with the signal basis, exploiting this sparsity allows the signal to be reconstructed from far fewer observations than Nyquist would suggest. The essential intuition here is that incoherence ensures that each observation is sensitive to many (ideally: all) coefficients within the basis function expansion; the principle of sparsity then allows us to assign the resulting information across the smallest number of coefficients possible.

In theory, imposing sparsity should require us to use a penalty term that counts the number of non-zero model coefficients: $\phi_m(\mathbf{m}) = \alpha^2\|\mathbf{m}\|_0$. However, this does not lead to a tractable computational problem. Instead, \citet{Donoho2006} has shown that it is sufficient to penalise the $L_1$ norm of the model vector, $\phi_m(\mathbf{m}) = \alpha^2\|\mathbf{m}\|_1=\alpha^2\sum_i |m_i|$. This can be implemented using a variety of algorithms, including quadratic programming techniques and the Lasso \citep{Tibshirani1996}. Costs are markedly higher than for $L_2$-based penalty functions, but remain tolerable.

Sparsity-promoting algorithms have significant potential: they open up new paradigms for data collection, offering the opportunity to substantially reduce the burden of storing, transmitting and handling datasets. The success of compressed sensing also suggests that the data misfit $\phi_d(\mathbf{d},\mathbf{g}(m))$ may be accurately estimated using only a small number of randomly-chosen samples: for certain classes of forward model, this may offer a route to substantially-reduced computational costs. 
Again, work is ongoing to explore the variety of ways in which concepts of sparsity can be applied and exploited within the context of geophysical inversion \citep[e.g][]{Herrmann2009,Wang2011,Simons2011a,Bianco2018,Muir2021}.

\subsection{Non-Euclidean Data Metrics}
A common challenge for gradient-based methods is convergence to a local---rather than global---minimum. This situation is difficult to identify or robustly avoid, since doing so would require knowledge of the global behaviour of the forward model. In this context, a particular downside to the use of a Euclidean data norm is that it treats each element of the data vector (i.e., each individual digitised data point) independently. For geophysical datasets, this is often undesirable: the spatial and temporal relationships connecting distinct data points are physically-meaningful, and a model that misplaces a data feature (such as a seismic arrival) in time or space is often preferable to one that fails to predict it at all. This problem is particularly familiar in waveform-fitting tasks, where the Euclidean norm is unduly sensitive to any phase differences between data and synthetics. From an optimisation perspective, this can manifest as `cycle-skipping', where waveforms end up mis-aligned by one or more complete periods. 

As a result, there is interest---and perhaps significant value---in exploring alternative metrics for quantifying the agreement between real and observed data sets. A particular focus of current research is measures built upon the theory of Optimal Transport \citep[e.g.][]{Ambrosio2003,Santambrogio2015}. This focusses on quantifying the `work' (appropriately defined) required to transform one object into another, and the most efficient path between the two states. In particular, the p-Wasserstein distance between two densities, $f(x)$ and $g(x)$ may be defined
\begin{equation}
W_p(f,g) = \left[\infimum_{T\in\mathcal{T}} \int c(x,T(x))^p f(x)\,\mathrm{d}x\right]^{1/p}
\end{equation}
where $\mathcal{T}$ is the set of all `transport plans' $T(x)$ that satisfy
\begin{equation}
f(x) = g(T(x))|\nabla T(x)|
\end{equation}
and $c(x,y)$ is a measure of the distance between points $x$ and $y$. The resulting metric provides a much more intuitive measure of the difference between two datasets, and perhaps offers a principled route to combining information from multiple distinct data types (sometimes known as `joint inversion').

Pioneered in geophysics by \citet{Engquist2014}, this has subsequently been employed for numerous studies, including the work of \citet{Metivier2016,Metivier2016a,Metivier2016b,Metivier2016c} and others \citep[e.g.][]{Huang2019,He2019,Hedjazian2019}. However, numerous challenges remain to be fully-overcome. Since Optimal Transport is conceived around density functions---which are inherently positive---signed datasets such as waveforms require special treatment. In addition, since computing the Wasserstein distance between two functions is itself an optimisation problem, there are practical challenges associated with employing it in large-scale inversion problems, and these are the focus of current work.

\section{Ensemble-Based Methods}
\begin{figure}
\includegraphics[width=\textwidth]{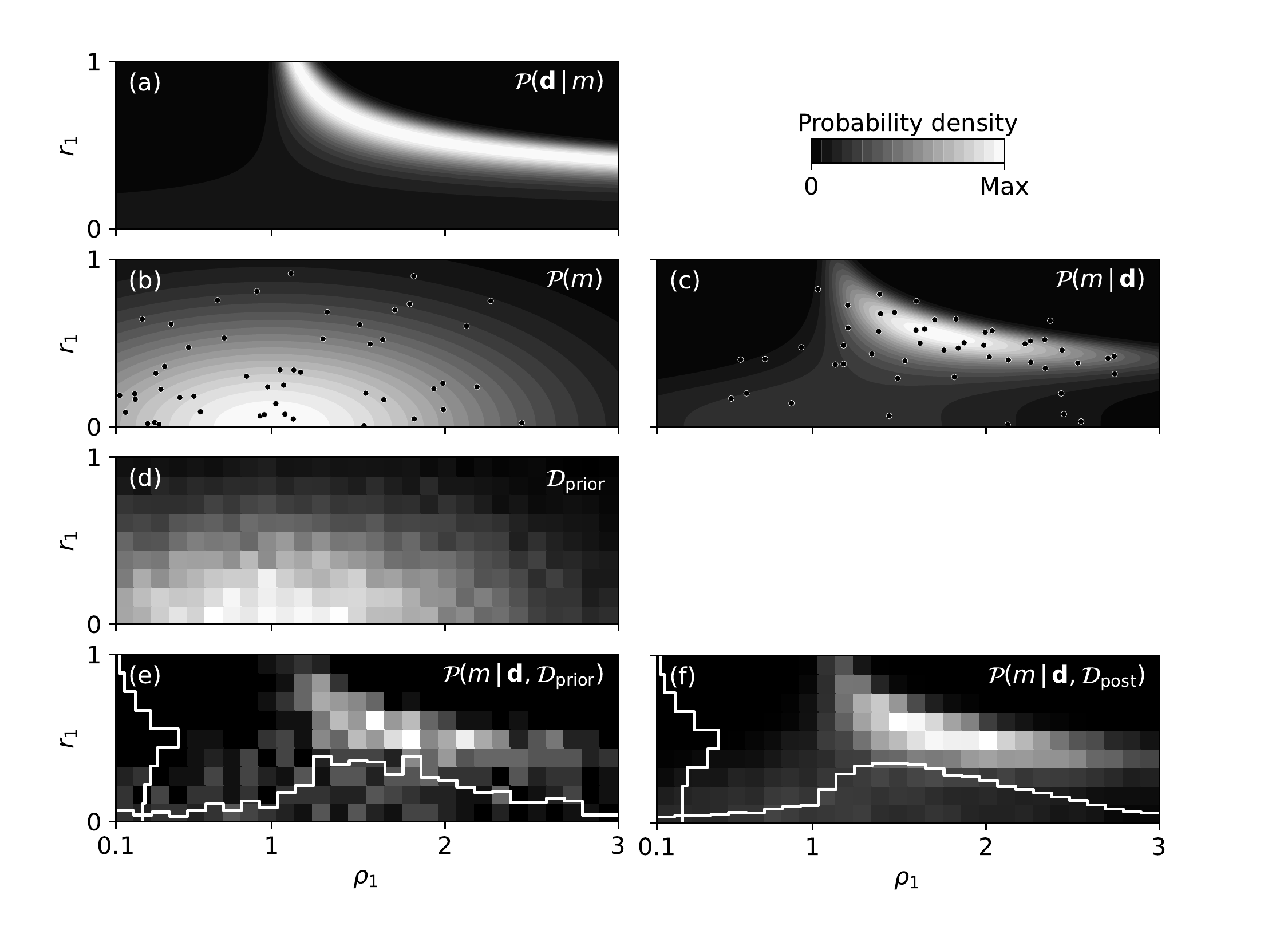}
\caption{\label{fig:bayes}Bayesian analysis for the simple inverse problem introduced in Fig.~\ref{fig:deterministic}. (a) The likelihood, $\prob{P}(\mathbf{d}\,|\,m)$, quantifies the extent to which any given choice of model can explain the data. (b) The prior distribution, $\prob{P}(m)$, encapsulates our beliefs \emph{before} observing any data, and can be `sampled' to generate a collection of candidate models ($\mathcal{D}_\mathrm{prior}$; dots; 50 shown). (c) The posterior distribution, $\prob{P}(m\,|\,\mathbf{d})$ combines prior and likelihood (eq.~\ref{eq:bayes}) to encapsulate our state of knowledge \emph{after} taking account of the data. In realistic problems visualising the posterior is intractable, but we can generate samples from it ($\mathcal{D}_\mathrm{post}$; 50 shown). (d) We can evaluate the forward model $g(m)$ for each example within an ensemble of prior samples, and additionally simulate the effects of noise processes. This can be completed without reference to any data. The information can be stored in many forms, including as a machine learning model. (e) Once data becomes available, this information can be queried to identify regions of parameter space that may explain observations---see Section \ref{sec:priorsamp}. This provides an approximation to the posterior; we additionally show 1-D marginals for each model parameter. (f) A similarly-sized set of posterior samples provides a much better approximation to the true posterior, as it is targeted towards explaining one specific set of observations---see Section \ref{sec:postsamp}. However, computational costs may be prohibitive for some applications.}
\end{figure}

We now switch focus, and consider the second fundamental approach to geophysical inversion: instead of seeking a single model that explains the data, we now aim to characterise the collection, or ensemble, of models that are compatible with observations. Clearly, this has potential to be more informative, providing insight into uncertainties and tradeoffs; however, it also brings new challenges. Computational costs may be high, and interpretation and decision-making may be complicated without the (illusion of) certainty promised by single-model strategies.

There are many different ways in which one might frame an ensemble-based inversion strategy: at the simplest, one might adapt the grid-search strategy of Section \ref{sec:single} so that the `ensemble' is the set of all grid nodes for which $\phi(m)$ is below some threshold. This approach, with models generated randomly rather than on a grid, underpinned some of the earliest ensemble-based studies in geophysics \citep[e.g.][]{Press1970,Anderssen1972,Worthington1972}. However, it is not particularly convenient from a computational perspective, since such an ensemble has little structure that can be exploited for efficiency or ease of analysis. Techniques exist that seek to address this \citep[e.g.][]{Sambridge1998} but the most common strategy is to adopt a probabilistic---and typically Bayesian---perspective. This involves a subtle, but important, change of philosophy: rather than seeking to determine the Earth structure directly, Bayesian inversion aims to quantify our state of knowledge (or `degree of belief') about that structure \citep[for more discussion, see, e.g.,][]{Scales1997}. 

The hallmark of Bayesian methods is that the posterior distribution---$\prob{P}(m\,|\,\mathbf{d})$, the probability of a model $m$ given the observations $\mathbf{d}$---is obtained by taking the prior distribution ($\prob{P}(m)$, our state of knowledge before making any observations), and weighting it by the likelihood, $\prob{P}(\mathbf{d}\,|\,m)$, which encapsulates the extent to which the data support any given model (see Fig.~\ref{fig:bayes}a--c). When normalised to give a valid probability distribution, we obtain 
\begin{equation}
\prob{P}(m\,|\,\mathbf{d}) = \frac{\prob{P}(\mathbf d\,|\,m) \,\prob{P}(m)}{\prob{P}(\mathbf d)}\label{eq:bayes}
\end{equation}
which is well-known as \indexterm{Bayes' Theorem} \citep{Bayes1763}.  We take this opportunity to remark that whereas a misfit function may be chosen in rather \emph{ad hoc} fashion to exhibit whatever sensitivity is desired, a likelihood has inherent meaning as `the probability that the observations arose from a given model', and ought to be defined by reference to the expected noise characteristics of the data. We also highlight the work of \citet{Allmaras2013}, which provides a comprehensive but accessible account of the practical application of Bayes' Theorem to an experimental inference problem. However, it is usually challenging to employ eq.~(\ref{eq:bayes}) directly, since evaluating the `evidence', $\prob{P}(\mathbf{d})$, requires an integral over the space of all allowable models, $\mathcal{M}$,
\begin{equation}
\prob{P}(\mathbf{d}) = \int_\mathcal{M} \prob{P}(\mathbf{d}\,|\,m)\prob{P}(m)\,\mathrm{d}m
\end{equation}
which is not computationally tractable for arbitrary large-scale problems. Instead, most Bayesian studies either make additional assumptions that enable analytic or semi-analytic evaluation of the evidence, or they exploit the fact that the ratio $\prob{P}(m_A\,|\,\mathbf{d})/\prob{P}(m_B\,|\,\mathbf{d})$ can be evaluated without knowledge of the evidence to obtain information about the \emph{relative} probability of different models.
\subsection{Bayesian \indexterm{Least Squares}}
The choice of prior is central to the success of any Bayesian approach---and also lies at the heart of many controversies and interpretational challenges, largely due to the impossibility of representing the state of no information \citep[e.g.][]{Backus1988}. It is therefore apparent that within a Bayesian framework all inference is considered relative to a known prior. In principle, the prior should be chosen based on a careful consideration of what is known about the problem of interest; in practice, this is often tempered by computational pragmatism, and a distribution with useful analytic properties is adopted.
\subsubsection{Gaussian Process Priors}\label{sec:gp}
A convenient choice when dealing with an unknown model function $m(\mathbf{x,t})$, is a \indexterm{Gaussian Process} prior,
\begin{equation}
m(\mathbf{x},t) \sim \mathcal{GP}\left(\mu(\mathbf{x},t), k(\mathbf{x},t,\mathbf{x^\prime},t^\prime) \right)\label{eq:gpprior}
\end{equation}
This is essentially the extension of the familiar normal distribution into function space, with our knowledge at any given point, $(\mathbf{x},t)$, quantified by a mean $\mu(\mathbf{x},t)$ and standard deviation $k(\mathbf{x},t;\mathbf{x},t)^{1/2}$; however, the covariance function $k$ also quantifies our knowledge (or assumptions) about the expected covariances if $m$ were to be measured at multiple distinct points. A comprehensive introduction to the theory of Gaussian Processes may be found in, e.g., \citet{Rasmussen2006}.

In some geophysical problems, the data-model relationship is---or can usefully be approximated as---linear (see also Section \ref{sec:euclid}), and so can be expressed in the form
\begin{equation}
d_i = \int_0^T\!\!\int_\mathcal{X} q_i(\mathbf{x},t) m(\mathbf{x},t)\,\mathrm{d}\mathbf{x}\,\mathrm{d}t\label{eq:datakernel}
\end{equation}
where $q_i(\mathbf{x},t)$ is some `data kernel',  and where $\mathcal{X}$ represents the domain upon which the model is defined. Moreover, we assume that the noise process represented by $\mathbf{C_d}$ is explicitly Gaussian. These assumptions permit analytic evaluation of the evidence, and the posterior distribution can be written in the form \citep{Valentine2020}
\begin{equation}
\tilde m(\mathbf{x},t) \sim \mathcal{GP}\left(\tilde \mu(\mathbf{x},t), \tilde k(\mathbf{x},t;\mathbf{x^\prime},t^\prime) \right)
\end{equation}
where we use a tilde to denote a posterior quantity, and where
\begin{align}
\tilde\mu(\mathbf{x},t)& = \mu(\mathbf{x},t) + \sum_{ij} w_i(\mathbf{x},t) \left[\left(\mathbf{{W}+C_d}\right)^\mathbf{-1}\right]_{ij}(d_j-{\omega}_j)\\
\tilde k(\mathbf{x},t;\mathbf{x^\prime},t^\prime)&=k(\mathbf{x},t;\mathbf{x^\prime},t^\prime) - \sum_{ij} w_i(\mathbf{x},t)\left[\left(\mathbf{{W}+C_d}\right)^\mathbf{-1}\right]_{ij}w_j(\mathbf{x^\prime},t^\prime)
\end{align}
with
\begin{align}
w_i(\mathbf{x},t)& = \int_0^T\!\!\int_\mathcal{X} k(\mathbf{x},t;\mathbf{x^\prime},t^\prime) q_i(\mathbf{x^\prime},t^\prime) \,\mathrm{d}\mathbf{x^\prime}\,\mathrm{d}t^\prime\\
W_{ij} &=  \int_0^T\!\!\int_0^T\!\!\int\!\!\!\!\int_{\mathcal{X}^2} q_i(\mathbf{x},t) k(\mathbf{x},t;\mathbf{x^\prime},t^\prime) q_j(\mathbf{x^\prime},t^\prime)\,\mathrm{d}\mathbf{x}\,\mathrm{d}\mathbf{x^\prime}\,\mathrm{d}t\,\mathrm{d}t^\prime\\
\omega_i &=\int_0^T\!\!\int_\mathcal{X} \mu(\mathbf{x},t)q_i(\mathbf{x},t)\,\mathrm{d}\mathbf{x}\,\mathrm{d}t\label{eq:omega}
\end{align}
This approach has formed the basis for a variety of geophysical studies \citep[e.g.][]{Tarantola1984,Montagner1990,Montagner1991,Valentine2020b} and has the attractive property that the inference problem is posed directly in a function space, avoiding some of the difficulties associated with discretization (such as spectral leakage).
\subsubsection{Discretised Form}
Nevertheless, if one chooses to introduce a finite set of basis functions, as in eq.~(\ref{eq:basis}), it is possible to express eqs.~(\ref{eq:gpprior}--\ref{eq:omega}) in discretized form \citep[for full discussion, see][]{Valentine2020a}. The prior distribution on the expansion coefficients becomes
\begin{align}
\mathbf{m}\sim\mathcal{N}\left(\mathbf{m_p},\mathbf{C_m}\right)
\end{align}
and the linear data-model relationship is expressed in the form $\mathbf{g(m)}=\mathbf{Gm}$. The posterior distribution may be written in a variety of forms, including
\begin{equation}
\mathbf{m} \sim \mathcal{N}(\mathbf{\tilde{m}},\mathbf{\tilde{C}_m})\label{eq:tv}
\end{equation}
where
\begin{align}
\mathbf{\tilde{m}} &=\mathbf{m_p} + \left(\mathbf{G^TC_d^{-1}G+C_m^{-1}}\right)^\mathbf{-1}\mathbf{G^TC_d^{-1}(d-Gm_p)}\\
\mathbf{\tilde{C}_m}&=\left(\mathbf{G^TC_d^{-1}G+C_m^{-1}}\right)^\mathbf{-1}
\end{align}
This well-known result, found in \citet{Tarantola1982}, has formed the basis of much work in geophysics. The expression for $\mathbf{\tilde{m}}$ is also often applied in non-Bayesian guise---compare with the discussion in Section \ref{sec:euclid}---with the prior covariance matrix $\mathbf{C_m}$ regarded as a generic `regularisation matrix' without probabilistic interpretation.
\subsection{\indexterm{Prior Sampling}}\label{sec:priorsamp}
The results of the previous section are built upon assumptions that our prior knowledge is Gaussian and the forward model is linear. This is computationally convenient, but will rarely be an accurate representation of the true state of affairs. Unfortunately, more general assumptions tend not to support analytic expressions for the posterior, and hence it becomes necessary to adopt `sampling-based methods'. These rely on evaluating the forward problem for a large number of models, in order to accumulate information about the relationship between model and data. Various strategies exist, which can be characterised by the manner in which sampling is performed.

The first group of strategies are those where candidate models are generated according to the prior distribution, and predicted data (potentially including simulated `noise') is computed for each. This provides a set of samples 
\begin{equation}
\mathcal{D}_\mathrm{prior} =\left\{ \left(\mathbf{m}_i, \mathbf{g}(\mathbf{m}_i)\right), \quad i=1,\ldots,N\right\}
\end{equation}
which may then be interpolated as necessary to address inversion questions (see Fig.~\ref{fig:bayes}d--e). This family of approaches is known as `prior sampling'  \citep{Kaufl2016a}, with different examples characterised by differing approaches to interpolation. Many of the recent studies that exploit machine learning to perform inversion may be seen within the prior sampling framework, although not all are explicitly Bayesian in design.

\subsubsection{Mixture Density Networks}\label{sec:mdn}
If we \emph{do} take a Bayesian approach, then we may note that the density of samples within $\mathcal{D}_\mathrm{prior}$ approximates---by construction---the joint probability density, $\prob{P}(\mathbf{m},\mathbf{d})$.  If we can fit an appropriate parametric density function to the samples, it is then straightforward to interpolate to obtain the conditional density $\prob{P}(\mathbf{m}\,|\,\mathbf{d})$ corresponding to observations (which we recognise to be the posterior distribution). One currently-popular way to achieve this is to employ Mixture Density Networks \citep[MDNs; ][]{Bishop1995}, which involve an assumption that the conditional distribution can be written as a Gaussian Mixture Model (GMM),
\begin{equation}
\prob{P}(m_j\,|\,\mathbf{d}) \approx \sum_{k=1}^K \frac{w_k(\mathbf{d})}{\sqrt{2\pi \sigma_k^2(\mathbf{d})}} \exp\left(-\frac{\left(m_j - \mu_k(\mathbf{d})\right)^2}{2 \sigma_k^2(\mathbf{d})}\right)\label{eq:mdn}
\end{equation}
where the weights $w_k$ (which are subject to an additional constraint, $\sum_k w_k =1$), means $\mu_k$ and standard deviations $\sigma_k$ that define the GMM are assumed to be functions of the data. These relationships may in turn be represented by a neural network. The set of prior samples, $\mathcal{D}_\mathrm{prior}$, is then used to optimise the neural network parameters, such that the expected value
\begin{equation}
\mathbb{E}_{\mathcal{D}_\mathrm{prior}}\left\{\prob{P}(m_j\,|\,\mathbf{d})\right\} = \frac{1}{N}\sum_{i=1}^N \prob{P}\left(\left[\mathbf{m_i}\right]_j\,|\,\mathbf{g}\left(\mathbf{m_i}\right)\right)
\end{equation}
is maximised. This approach has been applied to a variety of geophysical problems, including structural studies at global \citep[e.g.][]{Meier2007, Wit2014} and local \citep[e.g.][]{Earp2020,Mosher2021} scales, seismic source characterisation \citep{Kaufl2014}, and mineral physics \citep[e.g.][]{Rijal2021}. 
\subsubsection{Challenges and Opportunities}
The principal downside to prior sampling---discussed in detail by \citet{Kaufl2016a} in the context of MDNs, but applicable more broadly---is the fact that only a few of the samples within $\mathcal{D}_\mathrm{prior}$ will provide useful information about any given set of observations. In realistic problems, the range of models encompassed by the prior is large in comparison to the range encompassed by the posterior, and much computational effort is expended on generating predictions that turn out to have little similarity to observations. This is exacerbated by issues associated with the `curse of dimensionality', motivating the common choice (implicit in our notation for eq.~\ref{eq:mdn}) to use prior sampling to infer low- or uni-dimensional marginal distributions rather than the full posterior. Overall, the consequence is that prior sampling tends to yield rather broad posteriors, representing `our state of knowledge in the light of the simulations we have performed', rather than `the most we can hope to learn from the available data'. We also emphasise that results are wholly dependent on the choice of prior, and will be meaningless if this does not encompass the real earth system. This is perhaps obvious in an explicitly Bayesian context, but may be lost when studies are framed primarily from the perspective of machine learning.

The great benefit of prior sampling is that nearly all of the computational costs are incurred \emph{before} any knowledge of observed data is required. As a result, it may be effective in situations where it is desirable to obtain results as rapidly or cheaply as possible following data collection---e.g. to enable expensive numerical wave propagation simulations to be employed for earthquake early warning \citep{Kaufl2016}. We note parallels here to the use of scenario-matching approaches in the field of tsunami early warning \citep[e.g][]{Steinmetz2010}. It is also well-suited to applications where the same fundamental inverse problem must be solved many times for distinct datasets, perhaps representing observations repeated over time, or at many localities throughout at a region.

Prior sampling may also be effective in settings requiring what we term `indirect' inference, where the primary goal is to understand some quantity derived from the model, rather than the model itself. For example, in an earthquake early warning setting, one might seek to determine seismic source information with a view to then using this to predict tsunami run-up, or the peak ground acceleration at critical infrastructure sites \citep{Kaufl2015a}. In a prior sampling setting, one may augment $\mathcal{D}_\mathrm{prior}$ to incorporate a diverse suite of predictions, $\mathcal{D}_\mathrm{prior} = \left\{\left(\mathbf{m}_i,\mathbf{g_1}(\mathbf{m}_i),\mathbf{g_2}(\mathbf{m}_i),\ldots\right),\quad i=1,\ldots,N\right\}$, and then employ some interpolation framework to use observations of the process associated with (say) $\mathbf{g_1}$ to make inferences about $\mathbf{g_2}$. From a Bayesian perspective, this can be seen as a process of marginalisation over the model parameters themselves. 

\subsection{Posterior Sampling}\label{sec:postsamp}
As an alternative to prior sampling, one may set out to generate a suite of samples, $\mathcal{D}_\mathrm{post}$, distributed according to the posterior (see Fig.~\ref{fig:bayes}f). Again, there are a variety of ways this can be achieved---for example, a simple (but inefficient) approach might involve rejection sampling. More commonly, \indexterm{Markov chain Monte Carlo} (McMC) methods are employed, with the posterior forming the equilibrium distribution of a random walk.  
Encompassed within the term McMC lie a broad swathe of algorithms, of which the Metropolis-Hastings is probably most familiar, and the field is continually the subject of much development. We do not attempt to survey these advances, but instead direct the reader to one of the many recent reviews or tutorials on the topic \citep[e.g.][]{Brooks2011,Hogg2018,Luengo2020}.
 
As set out in \citet{Kaufl2016a}, prior and posterior sampling procedures generate identical results in the theoretical limit. However, in practical settings they are suited to different classes of problems. Posterior sampling approaches are directed towards explaining a specific dataset: this allows computational resources to be targeted towards learning the specifics of the problem at hand, but prevents expensive simulations from being `recycled' in conjunction with other datasets. It should also be noted that the `solution' obtained via posterior sampling takes the form of an ensemble of discrete samples. This can be challenging to store, represent, and interrogate in a meaningful way: many studies resort to reducing the ensemble to a single maximum-likelihood or mean model, and perhaps some statistics about the (co)variances associated with different parameters, and thereby neglect much of the power of McMC methods. Effective solutions to this issue may be somewhat problem-dependent, but remain the focus of much work.
 
\subsubsection{Improving Acceptance Ratios}
Generation of ensembles of posterior samples is inherently wasteful: by definition, one does not know in advance where samples should be placed, and hence for every `useful' sample, a large numbers of candidate models must be tested (i.e., we must evaluate the forward problem) and rejected. This is exacerbated by requirements for `burn-in' (so that the chain is independent of the arbitrary starting point) and `chain thinning' (to reduce correlations between consecutive samples), which also cause substantial numbers of samples to be discarded. Much effort is therefore expended on developing a variety of strategies to improve `acceptance ratios' (i.e., the proportion of all tested models that end up retained within the final ensemble).

One route forward involves improving the `proposal distribution', i.e. the manner in which samples are generated for testing. Ideally, we wish to make the proposal distribution as close as possible to the posterior, so that nearly all samples may be retained. Of course, the difficulty in doing so is that the posterior is not known in advance. An avenue currently attracting considerable interest is Hamiltonian McMC (HMC) methods, which exploit analogies with Hamiltonian dynamics to guide the random walk process towards `acceptable' samples \citep[see, e.g.][]{Neal2011,Betancourt2017}. In order to do so, HMC methods require, and exploit, knowledge of the gradient of the likelihood with respect to the model parameters at each sampling point. This provides additional information about the underlying physical problem, enabling extrapolation away from the sample point, and the identification of `useful' directions for exploration. To apply this idea, we must be able to compute the required gradients efficiently; early applications in geophysics have included seismic exploration and full-waveform inversion \citep[e.g.][]{Sen2017,Fichtner2019,Aleardi2020}.

In many cases, the fundamental physical problem of eq.~(\ref{eq:genfwd}) is amenable to implementation (eq.~\ref{eq:fwdprob}) in a variety of ways, depending on the assumptions made ($\mathcal{B}$). Usually, simplified assumptions lead to implementations with lower computational costs, at the expense of introducing systematic biases into predictions. Recently, \citet{Koshkholgh2021} has exploited this to accelerate McMC sampling, by using a low-cost physical approximation to help define a proposal distribution. Likelihood evaluations continue to rely on a more complex physical model, so that accuracy is preserved within the solution to the inverse problem---but the physically-motivated proposal distribution improves the acceptance rate and reduces overall computational costs. This is an attractive strategy, and seems likely to underpin future theoretical developments.

\subsubsection{Trans-Dimensional Inference}
In practice, McMC studies typically assume a discretised model, expressed relative to some set of basis functions in as in eq.~(\ref{eq:basis}), and the choice of basis functions is influential in determining the characteristics of the solution. In particular, the number of terms in the basis function expansion typically governs the flexibility of the solution, and the scale-lengths that can be represented. However, it also governs the dimension of the search space: as the number of free parameters in the model grows, so does the complexity (and hence computational cost) of the Monte Carlo procedure. Trans-dimensional approaches arise as an attempt to strike a balance between these two competing considerations: both basis set and expansion coefficients are allowed to evolve during the random walk process \citep{Green1995,Sambridge2006,Bodin2009,Sambridge2012}.

The trans-dimensional idea has been applied to a wide variety of geoscience problems, including source \citep[e.g][]{Dettmer2014} and structural \citep[][]{Burdick2017,Galetti2017,Guo2020} studies using seismic data, in geomagnetism \citep[][]{Livermore2018} and in hydrology \citep{Enemark2019}. It can be particularly effective in settings where basis functions form a natural hierarchy of scale lengths, such as with wavelets and spherical harmonics, although keeping track of information creates computational challenges \citep{Hawkins2015}. We note that model complexity is not confined only to length-scales: one can also employ a trans-dimensional approach to the physical theory, perhaps to assess whether mechanisms such as anisotropy are truly mandated by available data. The approach can also be employed to identify change-points or discontinuities within a function \citep[e.g][]{Gallagher2011}, and used in combination with other techniques such as {Gaussian Processes} \citep{Ray2019,Ray2021}.

\subsection{Variational Methods}\label{sec:variational}
One of the drawbacks of posterior sampling is the fact that the sampling procedure must achieve two purposes: it not only `discovers' the form of the posterior distribution, but also acts as our mechanism for representing the solution (which takes the form of a collection of appropriately-distributed samples). Large numbers of samples are often required to ensure stable statistics and `convincing' figures, even if the underlying problem itself is rather simple. To address this, one may introduce a parametric representation of the posterior distribution, and frame the inference task as determination of the optimal values for the free parameters---much as with the Mixture Density Network (section \ref{sec:mdn}). This approach, often known as \indexterm{Variational Inference} \citep[e.g.][]{Blei2017}, transforms inference for ensembles into an optimisation problem, and offers potentially-large efficiency gains.

We sketch the basic concept here, noting that a galaxy of subtly-different strategies can be found in recent literature \citep[see, e.g.][for a review]{Zhang2019}. As usual, our goal is to determine the posterior distribution, $\prob{P}(m\,|\,\mathbf{d})$. To approximate this, we introduce a distribution function $\prob{Q}(m\,|\,\boldsymbol{\theta})$ that has known form, parameterised by some set of variables $\boldsymbol{\theta}$---for example, we might decide that $\prob{Q}$ should be a Gaussian mixture model, in which case $\boldsymbol\theta$ would encapsulate the weights, means and variances for each mixture component. Our basic goal is then to optimize the parameters $\boldsymbol\theta$ such that $\prob{Q}(m\,|\,\boldsymbol{\theta})\approx \prob{P}(m\,|\,\mathbf{d})$.

To make this meaningful, we must---much as in section \ref{sec:single}---first define some measure of the difference between the two distributions. In Variational Inference, the usual choice is the Kullback-Leibler divergence \citep{Kullback1951},
\begin{align}
D_\mathrm{KL}(\prob{Q}\,\|\,\prob{P})& = \int \prob{Q}(m\,|\,\boldsymbol\theta) \log \frac{\prob{Q}(m\,|\,\boldsymbol\theta)}{\prob{P}(m\,|\,\mathbf{d})}\,\mathrm{d}m\nonumber\\
&=\mathbb{E}_{\prob{Q}(m\,|\,\boldsymbol\theta)}\left\{\log \frac{\prob{Q}(m\,|\,\boldsymbol\theta)}{\prob{P}(m\,|\,\mathbf{d})}\right\}
\end{align}
where the notation $\mathbb{E}_{\prob{Q}(m)}\{f(m)\}$ signifies `the expected value of $f(m)$ when $m$ is distributed according to $\prob{Q}$'.
Exploiting the properties of logarithms, and applying Bayes' Theorem, we can rewrite this in the form
\begin{equation}\label{eq:dkl}
D_\mathrm{KL}(\prob{Q}\,\|\,\prob{P}) =\log \prob{P}(\mathbf{d}) +\mathbb{E}_{\prob{Q}(m\,|\,\boldsymbol{\theta})}\{\log\prob{Q}(m\,|\,\boldsymbol{\theta})-\log \prob{P}(\mathbf{d}\,|\,m)-\log\prob{P}(m)\}
\end{equation}
where $\prob{P}(\mathbf{d})$ has been moved outside the expectation since it is independent of $m$. While this quantity is unknown, it is also constant---and so can be neglected from the perspective of determining the value of $\boldsymbol\theta$ at which $D_{\mathrm{KL}}$ is minimised. The quantity $\prob{P}(\mathbf{d})-D_\mathrm{KL}(\prob{Q}\,\|\,\prob{P})$ is known as the `evidence lower bound' (ELBO), and maximisation of this is equivalent to minimising the Kullback-Leibler divergence. Because the variational family $\prob{Q}(m\,|\,\boldsymbol\theta)$ has a known form, the ELBO can be evaluated, as can the derivatives ${\partial D_\mathrm{KL}}/{\partial \theta_i}$. Thus, it is conceptually straightforward to apply any gradient-based optimisation scheme to determine the parameters such that $\prob{Q}$ best approximates the posterior distribution.

\subsubsection{A Gaussian Approximation}
To illustrate this procedure, and to highlight connections to other approaches, we consider an inverse problem where: (i) the model is discretised, as in eq.~(\ref{eq:basis}), so that we seek an $M$-component model vector $\mathbf{m}$; the prior distribution on those model coefficients is Gaussian with mean $\mathbf{m_p}$ and covariance $\mathbf{C_m}$; and (iii) the likelihood takes the form $\prob{P}(\mathbf{m}\,|\,\mathbf{d}) = k \exp( -\frac{1}{2}\phi(\mathbf{m}))$ for some appropriate function $\phi$. We choose to assert that the solution can be approximated by a Gaussian of mean $\boldsymbol\mu$ and covariance matrix $\boldsymbol\Sigma$, and seek the optimal values of these quantities. Thus, we choose
\begin{equation}
\prob{Q}(\mathbf{m}\,|\,\boldsymbol{\mu},\boldsymbol{\Sigma}) = \frac{1}{(2\pi)^{M/2}(\det \boldsymbol\Sigma)^{1/2}}\exp\left\{-(\mathbf{m}-\boldsymbol{\mu})^\mathbf{T}\boldsymbol\Sigma^\mathbf{-1}(\mathbf{m}-\boldsymbol{\mu})\right\}\,\mathrm{.}
\end{equation}
To proceed, we need to determine the expectation of various functions of $\mathbf{m}$ under this distribution---and their gradients with respect to $\boldsymbol{\mu}$ and $\boldsymbol{\Sigma}$.

A number of useful analytical results and expressions can be found in \citet{Petersen2015}. It is straightforward to determine that
\begin{align}
\frac{\partial}{\partial \boldsymbol{\mu}}D_\mathrm{KL}(\prob{Q}\,\|\,\prob{P})&=\mathbf{C_m^{-1}}(\boldsymbol\mu - \mathbf{m_p}) + \frac{1}{2}\frac{\partial}{\partial\boldsymbol\mu}\mathbb{E}_\prob{Q}\left\{\phi(\mathbf{m})\right\}\label{eq:vigrad1}\\
\frac{\partial}{\partial \boldsymbol{\Sigma}}D_\mathrm{KL}(\prob{Q}\,\|\,\prob{P})&=-\frac{1}{2}\left(\boldsymbol\Sigma^\mathbf{-1} - \mathbf{C_m^{-1}}\right)+\frac{1}{2}\frac{\partial}{\partial\boldsymbol\Sigma}\mathbb{E}_\prob{Q}\left\{\phi(\mathbf{m})\right\}\label{eq:vigrad2}
\end{align}
These expressions can be used to drive an iterative optimisation procedure to determine the optimal variational parameters. In implementing this, the result
\begin{equation}
\frac{\partial}{\partial \theta_i} \mathbb{E}_{\prob{Q}(m\,|\,\boldsymbol\theta)}\left\{f[m]\right\} = \mathbb{E}_{\prob{Q}(m\,|\,\boldsymbol\theta)} \left\{f(m) \frac{\partial}{\partial \theta_i}\log\prob{Q}(m\,|\,\boldsymbol\theta)\right\}
\end{equation}
may be useful.

In the case where $\mathbf{g(m)}$ is (or is assumed to be) linear, and where the function $\phi$ is defined as the $L_2$ norm of the residuals, the expected value can be evaluated analytically. The misfit is quadratic in form, 
\begin{equation}
\phi(\mathbf{m}) = \mathbf{d^TC_d^{-1}d} - 2\mathbf{d^TC_d^{-1}Gm} + \mathbf{m^TG^TC_d^{-1}Gm}
\end{equation}
as in Section \ref{sec:gp}. Hence the expected value, given $\mathbf{m}$ is distributed according to the Gaussian $\prob{Q}$, can be determined, along with its derivatives
\begin{align}
\mathbb{E}_\prob{Q}\left\{\phi(\mathbf{m})\right\} &= -2\mathbf{d^TC_d^{-1}G}\boldsymbol{\mu} + \trace \left(\mathbf{G^TC_d^{-1}G\boldsymbol\Sigma} \right)+\boldsymbol{\mu}^\mathbf{T}\mathbf{G^TC_d^{-1}G}\boldsymbol{\mu}\\
\frac{\partial}{\partial\boldsymbol\mu}\mathbb{E}_\prob{Q}\left\{\phi(\mathbf{m})\right\}&=- 2\mathbf{G^TC_d^{-1}d}+2\mathbf{G^TC_d^{-1}G}\boldsymbol\mu \\
\frac{\partial}{\partial\boldsymbol\Sigma}\mathbb{E}_\prob{Q}\left\{\phi(\mathbf{m})\right\}&=\mathbf{G^TC_d^{-1}G}
\end{align}
Substituting these expressions into eqs.~(\ref{eq:vigrad1}--\ref{eq:vigrad2}), and solving for the $\boldsymbol\mu$ and $\boldsymbol\Sigma$ such that the gradients of $D_\mathrm{KL}$ are zero (as is required at a minimum), we find that the optimal distribution $\prob{Q}$ is identical to the posterior distribution obtained in eq.~(\ref{eq:tv}). This is unsurprising, since our underlying assumptions are also identical---but demonstrates the self-consistency of, and connections between, the different approaches. Again, we also highlight the similarity with the expressions obtained in Section \ref{sec:euclid}, although the underlying philosophy differs.
\subsubsection{Geophysical Applications}
Variational methods offer a promising route to flexible but tractable inference. As the preceding example illustrates, they provide opportunities to balance the (assumed) complexity and expressivity of the solution against computational costs. A number of recent studies have therefore explicitly sought to explore their potential in particular applications, including for earthquake hypocentre determination \citep{Smith2022}, seismic tomography \citep{Zhang2020,Siahkoohi2021,Zhao2022} and hydrogeology \citep{Ramgraber2021}. However, given the fairly broad ambit of variational inference, many past studies could also be seen as falling under this umbrella.

\subsection{\indexterm{Generative Models}}
Many of the methods discussed so far rely on strong assumptions about the form of prior and/or posterior distributions: we suppose that these belong to some relatively simple family, with properties that we can then exploit for efficient calculations. However, such assertions are typically justified by their convenience---perhaps aided by an appeal to the principle known as Occam's Razor---and not through any fundamental physical reasoning \citep[see, e.g.][]{Constable1987}. This is unsatisfactory, and may contribute substantial unquantifiable errors into solutions and their associated uncertainty estimates.

Recently, a number of techniques have emerged that allow representation of, and computation with, relatively general probability distributions. In broad terms, these are built upon the idea that arbitrarily complex probability distributions can be constructed via transformations of simpler distributions. This is familiar territory: whenever we need to generate normally-distributed random numbers, a technique such as the  Box-Muller transform \citep{Box1958} is applied to the uniformly-distributed output of a pseudo-random number generator. However, the versatility of such approaches is vastly increased in conjunction with the tools and techniques of modern machine learning.

This is an area that is currently the focus of rapid development; recent reviews include those of \citet{Bond-Taylor2022} and \citet{Ruthotto2021}. Clearly, the concept is closely-connected to the idea of variational inference, as discussed in Section \ref{sec:variational}. Several major techniques have emerged, including `generative adversarial networks' (GANs) \citep[e.g.][]{Goodfellow2014, Creswell2018}, `variational autoencoders' \citep{Kingma2014}, and `normalizing flows' \citep{Rezende2015,Kobyzev2021}. A variety of recent studies have explored diverse applications of these concepts within the context of geophysical inversion: examples include \citet{Mosser2020}, \citet{Lopez-Alvis2021}, \citet{Zhao2022} and \citet{Scheiter2022}. We have no doubt that this area will lead to influential developments, although the precise scope of these is not yet clear.

\section{Model Properties}
The third fundamental approach builds on the work of \citet{Backus1968} and \citet{Backus1970a,Backus1970b,Backus1970c}, and we sketch it briefly for completeness. For certain classes of problem, as in eq.~(\ref{eq:datakernel}), each of the observables $d_i$ can be regarded as representing an average of the model function weighted by some data kernel $q_i(\mathbf{x},t)$. It is then straightforward to write down a weighted sum of the observations,
\begin{equation}
D_{\boldsymbol\alpha} = \sum_i \alpha_i d_i 
=\int_0^T\!\!\!\int_\mathcal{X} Q(\boldsymbol{\alpha},\mathbf{x},t) m(\mathbf{x},t)\,\mathrm{d}\mathbf{x}\,\mathrm{d}t
\end{equation}
where $Q(\boldsymbol{\alpha},\mathbf{x},t) =\sum_i \alpha_i q_i(\mathbf{x},t)$, and $\boldsymbol\alpha = (\alpha_1,\alpha_2\ldots)$ represents some set of tunable weights. By adjusting these, one may vary the form of the averaging kernel $Q$, and frame a functional optimisation problem to determine the $\boldsymbol\alpha$ that brings $Q$ as close as possible to some desired form. In this way, the value of the average that is sought can be estimated as a linear combination of the observed data.

\indexterm{Backus-Gilbert theory} has an inherent honesty: it is data-led, with a focus on understanding what the available data can---or cannot---constrain within the system. On the other hand, this can be seen as a downside: it is not usually possible to use the results of a Backus-Gilbert--style analysis as the foundation for further simulations. Moreover, interpretation can be challenging in large-scale applications, as the `meaning' of each result must be considered in the light of the particular averaging kernel found. Perhaps for this reason---and because it is designed for strictly linear problems \citep[although we note the work of][]{Snieder1991}---the method is well-known but has found comparatively little use. Notable early examples include \citet{Green1975}, \citet{Chou1979} and \citet{Tanimoto1985,Tanimoto1986a}. More recently, it has been adopted by the helioseismology community \citep{Pijpers1992}, and applied to global tomography \citep{Zaroli2016}  and to constrain mantle discontinuities \citep{Lau2021}. Concepts from Backus-Gilbert theory are also sometimes used to support interpretation of models produced using other approaches: for example, \citet{Ritsema1999} presents Backus-Gilbert kernels to illustrate the resolution of a model obtained by least-squares inversion.

\section{Miscellanea}
The preceding sections have focussed on the range of different philosophies, and associated techniques, by which geophysical inversion can be framed. We now turn to consider some additional concepts and developments that are not themselves designed to solve inverse problems, but which can potentially be employed in conjunction with one or other of the approaches described above.

\subsection{Approximate Forward Models}
One of the major limiting factors in any geophysical inversion is computational cost. High-fidelity numerical models tend to be computationally-expensive, and costs may reach hundreds or even thousands of cpu-hours per simulation. In such cases, resource availability may severely constrain the number of simulations that may be performed, rendering certain approaches infeasible. There is therefore considerable potential value in any technique that may lower the burden of simulation.

\subsubsection{Surrogate Modelling}
One possible solution to this lies in `\indexterm{surrogate modelling}': using techniques of machine learning to mimic the behaviour of an expensive forward model, but at much lower computational cost. This is an idea that has its origins in engineering design \citep[see, e.g.,][]{Quiepo2005,Forrester2009}, and typically involves tuning the free parameters of a neural network or other approximator to match a database of examples obtained via expensive computations (or, indeed, physical experiments). The approximate function can then be interrogated to provide insights, or to serve as a drop-in replacement for the numerical code.
 
Although the term `surrogate modelling' only appears relatively recently in the geophysics literature, the underlying idea has a long history. For example, seismologists have long recognised that travel times of seismic arrivals from known sources can be interpolated, and the resulting travel-time curves used to assist in the location of new events \citep[e.g.][]{Jeffreys1940,Kennett1991,Nicholson2004a}. One may also regard the Neighbourhood Algorithm \citep{Sambridge1999,Sambridge1999a} within this framework: it uses computational geometry to assemble a surrogate approximation to evaluation of (typically) the likelihood for any given model. By employing and refining this within a Markov chain, it is possible to substantially reduce the computational costs of McMC-based inference. In doing so, we exploit the fact that the mapping from models to likelihood (a scalar quantity) is, typically, much simpler than the mapping from models to data. Closely-related is the field of `Bayesian optimization', which relies on a surrogate (often a Gaussian Process) to encapsulate incomplete knowledge of an objective function, and takes this uncertainty into account within the optimization procedure \citep[e.g.][]{Shahriari2016, Wang2016}.

Latterly, surrogate models (also known as emulators) have been explicitly adopted for geophysical studies. Similar to the Neighbourhood Algorithm, \citet{Chandra2020} employed a neural network-based surrogate to replace likelihood calculations within a landscape evolution model; on the other hand, \citet{Das2018} and \citet{Spurio-Mancini2021} both develop a surrogate that directly replaces a forward model and outputs synthetic seismograms. Other geophysical examples include modelling of climate and weather \citep[e.g.][]{Field2011,Castruccio2014}, and applications in hydrology \citep{Hussain2015} and planetary geophysics \citep[][]{Agarwal2020}.

\subsubsection{Physics-Informed Neural Networks}
A number of recent studies have also explored the concept  and applications of `physics-informed neural networks'  \citep[PINNs; see, e.g.][]{Raissi2019,Karniadakis2021}. As with surrogate models, these exploit machine learning techniques to provide a version of the forward model that has significantly lower computational cost than `conventional' implementations. However, whereas a surrogate is constructed using a suite of examples obtained by running the conventional model (at substantial expense), a PINN is directly trained to satisfy the physical constraints. Typically, this amounts to defining a neural network to represent the observable function, $u(\mathbf{x},t)$, and then employing a training procedure to  minimise the deviation from eq.~(\ref{eq:genfwd}). This is potentially a more efficient approach, and provides the researcher with greater oversight of the behaviour and limitations of the learned model. 

A number of recent examples may be found, particularly in the seismological literature. \citet{Moseley2020}, \citet{Song2021} and \citet{Smith2020} all use PINNs to solve problems related to the wave equation, with the latter underpinning the variational inference approach of \citet{Smith2022}. A range of potential applications in climate science and meteorology are discussed in \citet{Kashinath2021}, while \citet{He2021} consider hydrological problems. Again, it is clear that PINNs present a promising opportunity that is likely to bring substantial benefits for geophysics, but it is not yet clear how the field will evolve.

\subsubsection{Conventional Approximations}
Surrogate models and PINNs both rely on machine learning, and their `approximate' nature arises from this: they are constructed to give good average performance for a particular task, but there are few hard constraints on their accuracy in any specific case. In many geophysical problems, an alternate route exists, and has long been exploited: rather than seeking an approximate solution to a complex physical problem, we can use conventional methods to obtain an accurate solution for a simplified physical system (i.e., adopting a more restrictive set of assumptions, $\mathcal{A}\cup\mathcal{B}$). Thus, for example, seismic waves might be modelled under the assumption that propagation is only affected by structure in the great-circle plane between source and receiver \citep{Woodhouse1984} at far lower cost than (almost) physically-complete simulation \citep[e.g][]{Komatitsch2002}. Depending on circumstances, it may be beneficial to exploit a known approximation of this kind, where impacts can be understood and interpretations adjusted accordingly. We also highlight that it may be desirable to vary the level of approximation used for forward simulations within an inversion framework, using a fast approximate technique for initial characterisation, and increasing accuracy as solutions are approached. In the ideal case, one might envisage a forward model where the level of approximation is itself a tuneable parameter \citep[e.g. via the coupling band-width in a normal-mode--based solver,][]{Woodhouse1980}, enabling a smooth transition from simplified to complete modelling as a solution is approached.

\subsection{Computational Advances}
Modern geophysics is computationally-intensive, and---as we have seen---the feasibility of various inversion strategies is directly linked to the available resources. As such, computational developments are often important in driving the development and adoption of novel inference approaches. In particular, current progress leverages a number of technological advances that have been stimulated by the rapid growth of `machine learning' applications across society. This includes general-purpose computational libraries such as Tensorflow \citep{Abadi2016} and Pytorch \citep{Paszke2019}, along with more specialist tools such as Edward \citep{Tran2016}. A key feature of these libraries is native support for auto-differentiation, making it easy to exploit gradient-based optimisation strategies. This is an area that has previously been highlighted as ripe for exploitation in geophysics \citep{Sambridge2007}, although its use is not yet widespread. Another interesting development is the rise of packages such as FEniCS \citep{Logg2012}, which aim to automatically generate forward models from a statement of the relevant physical equations \citep[e.g.][]{Reuber2020}. This has the potential to greatly expand the range of problems that it is feasible to address.

\subsection{Novel Data--Novel Strategies}
An ongoing theme of geophysics is the growth in data quantity. This is often driven by concerted efforts to collect high-resolution datasets: examples include high-quality satellite gravity measurements \citep[e.g.][]{Kornfeld2019}, and systematic continental-scale surveys such as USArray \citep{Meltzer1999} or AusAEM \citep{Ley-Cooper2020}. Handling and processing such massive datasets has necessitated new tools and standards designed to enable easy exploitation of high-performance computing \citep[e.g.][]{Krischer2016,Hassan2020}. On the other hand, we have also seen exciting recent developments in planetary seismology, with the recent breakthrough analysis of Martian seismic data from the InSight mission \citep{Knapmeyer-Endrun2021, Khan2021,Stahler2021}. In this context, the available dataset is very limited: we must work with a single instrument, limited capacity for data transmission, and with data characteristics quite different from those of Earth. Undoubtedly techniques will need to develop accordingly.

Another driver for innovation in geophysical inversion is innovation in data collection.  Recent advances in sensor technology includes the growth of distributed acoustic sensing \citep[e.g.][]{Daley2013,Parker2014}, which uses fibre-optic cables to measure strain rates, and nodal seismic acquisition systems \citep{Dean2018}, which enable dense deployments of semi-autonomous instruments. Fully-exploiting these technologies within an inversion context will doubtless motivate a new generation of analysis techniques \citep[e.g.][]{Lythgoe2021,Muir2021}, and ongoing innovation in the field of geophysical inversion.

\section{Concluding Remarks}
Athanasius \citeauthor{Kircher1665} published his \emph{Mundus Subterraneus} in 1665, with his now-famous images of fiery chambers criss-crossing the Earth's interior to feed its volcanoes. What was his evidence for this structure? He acknowledges: `\emph{sive ea jam hoc modo, sive alio}'---`either like this, or something else'. As \citet{Waddell2006} writes, this
\begin{quote} ...makes very clear that Kircher was not interested in whether his images had managed to capture exactly the subterranean structure of the Earth. Such large and detailed copper engravings must have been extremely expensive to commission and print, suggesting that Kircher did believe them to be important. But their value lay in their ability to encourage speculation and consideration...\end{quote}
Some 350 years later, geophysical images are produced with more emphasis on rigour---but otherwise, perhaps little has changed.

In this chapter, we have sought to survey and summarise the state of the art of geophysical inversion, and to highlight some of the theoretical and conceptual connections between different approaches. As we hope is clear, the field continues to develop at pace: driven by the need to better-address geoscience questions; drawn on towards exciting horizons across mathematics, statistics and computation. In particular, the growth of machine learning has focussed much attention on techniques of regression, model-building and statistical inference, and the fruits of this have been evident throughout our discussion.  We have no doubt that geophysical inversion will continue to produce images and models that can inspire and stimulate geoscientists for many years to come.


\section*{Acknowledgements}
We are grateful to the many students, colleagues and collaborators who have contributed to our understanding of the topics discussed in this chapter. We also thank several colleagues, and an anonymous reviewer, for helpful comments and suggestions on a draft of this work. We acknowledge financial support from the CSIRO Future Science Platform in Deep Earth Imaging, and from the Australian Research Council under grant numbers DP180100040 and DP200100053.
\clearpage
\bibliographystyle{gji}
\bibliography{emergingdirections.bib}
\end{document}